\documentclass[a4paper,11pt]{article}
\usepackage{mrs2005,epsfig}
\usepackage{amsmath}
\usepackage{amsfonts}
\usepackage{graphicx}
\usepackage{graphics}
\usepackage{epsfig}
\usepackage{rotate}

\begin{document}

\title{ Studying the global dynamics of conservative dynamical systems
using the SALI chaos detection method}

\author{ T. Manos$^{\ast,\dag}$, Ch. Skokos$^{\ddag}$,
E. Athanassoula$^{\dag}$ and T. Bountis$^{\ast}$}{ } \affil{ $^{\ast}$
Center for Research and Applications of Nonlinear Systems (CRANS),
Department of Mathematics, University of Patras, GR--26500, Greece.\\
$^{\dag}$ Observatoire Astronomique de Marseille-Provence (OAMP), 2
Place Le Verrier, F--13248, Marseille, France.\\
$\ddag$ Astronomie et Syst\`{e}mes Dynamiques, Institut de
M\'{e}canique C\'{e}leste et de Calcul des Eph\'{e}m\'{e}rides
(IMCCE), Observatoire de Paris, 77 avenue Denfert-Rochereau,
F--75014,\\
 Paris, France.  }

\begin{abstract}
We use the Smaller ALignment Index (SALI) method of chaos detection,
to study the global dynamics of conservative dynamical systems
described by differential or difference equations. In particular, we
consider the well--known 2 and 4--dimensional symplectic standard map,
as well as an autonomous Hamiltonian system of 2 and 3 degrees of
freedom describing the motion of a star in models of barred
galaxies. The application of SALI helped us to compute rapidly and
accurately the percentage of regular and chaotic motion for particular
values of the parameters of these systems. We were also able to
perform a computationally efficient determination of the dependence of
these percentages on the variation of several parameters of the
studied models.\\

\vspace{0.2cm}

\end{abstract}

\section{Introduction}
\label{Intro}

The qualitative distinction between chaotic and regular motion in
symplectic maps and in systems of differential equations is a
fundamental problem of non--linear dynamics. This distinction is in
general, a non trivial task and it becomes more difficult as the
number of degrees of freedom increases. For this reason, over the
years, several methods distinguishing regular from chaotic motion in
conservative dynamical systems have been proposed and applied, with
varying degrees of success.

One of the most efficient methods of chaos detection is the
computation of the so--called \textbf{S}maller \textbf{AL}ignment
\textbf{I}ndex (\textbf{SALI}) which was introduced in \cite{sk:1} and
has already been applied successfully to several dynamical systems
$[1\mbox{--}18]$.
SALI has proved to be a fast and reliable method which can distinguish
between regular and chaotic motion rapidly, reliably and
accurately. These characteristics make this index perfectly suited for
the study of global dynamics of dynamical systems. For these reasons
we use the SALI to study the behavior of two distinct dynamical
systems: the well--known 2--dimensional (2D) standard map \cite{Ch79}
and its generalization to higher dimensions, as well as Hamiltonian
systems of 2 (2D) and 3 (3D) degrees of freedom, describing the motion
of stars in models of barred galaxies. In the present paper, we
present some preliminary results of our studies.

The paper is organized as follows: In Section \ref{SALI} we recall the
definition of SALI explaining also its behavior for regular and
chaotic orbits. In Section \ref{SM} we use SALI for computing the
fraction of chaotic orbits in the case of the standard map, while in
Section \ref{Galaxies} the results of an analogous study for
Hamiltonian systems of barred galaxies are presented. Finally, in
Section \ref{Conclusions} we present our conclusions.

\section{The Smaller Alignment Index (SALI)}
\label{SALI}

Let us consider a $l$ - dimensional phase space of a conservative
dynamical system, which could be a $2M$-dimensional symplectic map or
a Hamiltonian flow of $N$ degrees of freedom, with $l=2N$. We consider
also an orbit in this space with initial condition
$S(0)=(x_{1}(0),x_{2}(0),...,x_{l}(0))$ and two deviation vectors
$V_{1}(0)=(dx_{11}(0),dx_{12}(0),...,dx_{1l}(0))$ and
$V_{2}(0)=(dx_{21}(0),dx_{22}(0),...,dx_{2l}(0))$, from the initial
point $S(0)$.

For the computation of the SALI of a given orbit, one has to follow
the time evolution of the orbit itself and also of two deviation
vectors which initially point in two different directions. The
evolution of an orbit of a map $F$ is described by the discrete-time
equations of the map:
\begin{equation}\label{map_eq:1}
    S(n+1)=F(S(n)),
\end{equation}
where $S(n)=(x_{1}(n),x_{2}(n),...,x_{l}(n))$, represents the
orbit's coordinates at the $n$-th iteration. The evolution of the
deviation vectors $V_{1}(n),V_{2}(n)$, in this case, is given by the
equations of the tangent map:
\begin{equation}\label{tan_mao:1}
    V(n+1)=DF(S(n))\cdot V(n).
\end{equation}
In the case of Hamiltonian flows the evolution of an orbit is given
by Hamilton's equations of motion:
\begin{equation}\label{eq_motion:1}
    \frac{dS(t)}{dt}=F(S(t)),
\end{equation}
where $F$ is a set of $n$-functions $(F_{1},F_{2},...,F_{n})$, while
the corresponding evolution of the deviation vectors
$V_{1}(t),V_{2}(t)$, is given by the variational equations:
\begin{equation}\label{var_equa:1}
    \frac{dV(t)}{dt}=DF(S(t))\cdot V(t)\,\, .
\end{equation}
We note that in (\ref{tan_mao:1}) and (\ref{var_equa:1}) $DF$
denotes the Jacobian matrix of equations (\ref{map_eq:1}) and
(\ref{eq_motion:1}) respectively, evaluated at the points of the
orbit under study.

At every time step (or iteration) the two deviation vectors
$V_{1}(t)$ and $V_{2}(t)$ are normalized with norm equal to 1 and
the SALI is then computed as:
\begin{equation}\label{eq:SALI:2}
    SALI(t)= min
    \left\{\left\|\frac{V_{1}(t)}{\|V_{1}(t)\|}+\frac{V_{2}(t)}{\|V_{2}(t)\|}\right\|,\left\|\frac{V_{1}(t)}{\|V_{1}(t)\|}-\frac{V_{2}(t)}{\|V_{2}(t)\|}\right\|\right\},
\end{equation}
where $\| \cdot \|$ denotes the usual Euclidean norm and $t$ is the
continuous or discrete time.

SALI has a completely different behavior for regular and chaotic
orbits and this allows us to clearly distinguish between them.  In the
case of Hamiltonian flows or $2M$--dimensional symplectic maps with
$2M > 2$, the SALI fluctuates around a non-zero value for regular
orbits, while it tends exponentially to zero for chaotic orbits
\cite{sk:1,sk:3,sk:4}, following a rate which depends on the
difference of the two largest Lyapunov Exponents \cite{sk:5}. Thus, in
2D and 3D Hamiltonian systems the distinction between ordered and
chaotic motion is easily done. On the other hand, in the case of 2D
maps the SALI tends to zero both for regular and for chaotic orbits,
following however completely different time rates, which again allows
us to distinguish between the two cases \cite{sk:1}.

\section{Global dynamics of 2D and 4D standard map}
\label{SM}

As a simple 2D map which exhibits regular and chaotic behavior, we
consider the well--known standard map \cite{Ch79} in the form
\begin{equation}\label{sm:1}
  \begin{array}{ll}
    x_{n+1} = x_n + y_{n+1} & \hbox{} \\
    y_{n+1} = y_n + \frac{K}{2\pi}\sin(2\pi x_n) & \hbox{}
  \end{array}  \,\, (mod\,\,1)\,\, ,
\end{equation}
where $K$ is the so--called non--linear parameter of the system.

Before studying the global dynamics of map (\ref{sm:1}) let us examine
in more detail the behavior of SALI for regular and chaotic orbits of
a 2D map. In the case of a chaotic orbit, any two deviation vectors
will be aligned to the direction defined by the largest Lyapunov
exponent $L_1$, and consequently SALI tends to zero following an
exponential decay of the form $\mbox{SALI} \propto e^{-2L_1 n}$, with
$n$ being the number of iterations \cite{sk:5}.  In the case of
regular orbits any two deviation vectors tend to fall on the tangent
space of the torus on which the motion lies \cite{sk:1,
sk:3,SBA06}. For a 2D map this torus is an 1--dimensional invariant
curve, whose tangent space is also 1--dimensional and consequently any
two deviation vectors will become aligned. Thus, even in the case of
regular orbits in 2D maps the SALI tends to zero. This decay follows a
power law \cite{sk:1} having the form $\mbox{SALI} \propto 1/n^2$
\cite{SBA06}.

\begin{figure}
\includegraphics[height=0.3\textheight]{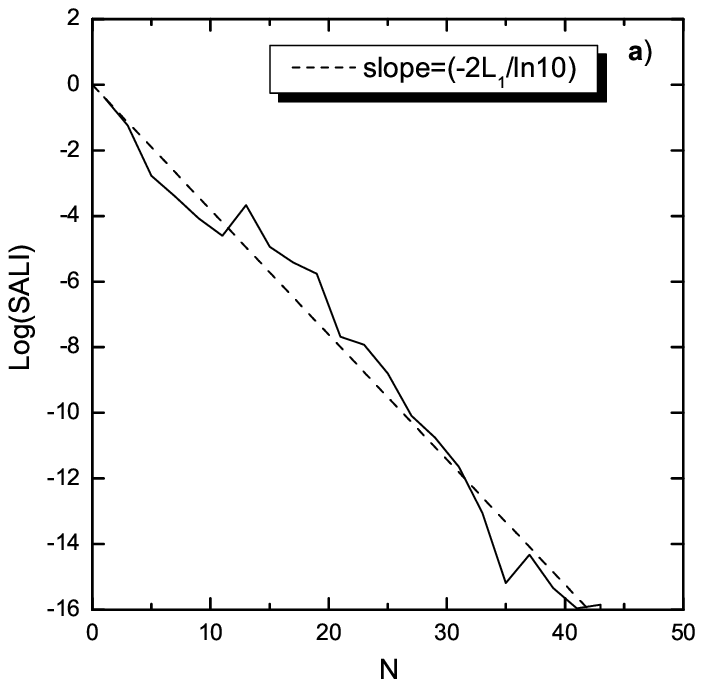}\hspace{0.2cm}
\includegraphics[height=0.3\textheight]{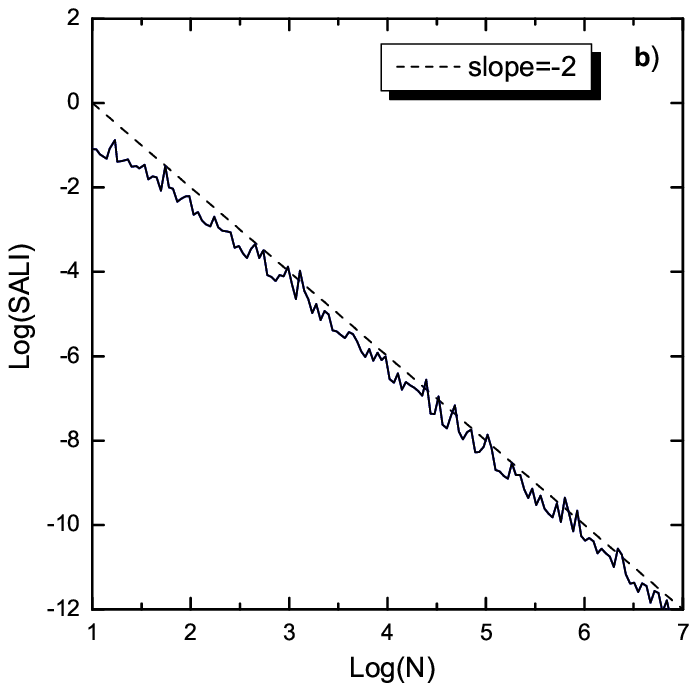}
\caption{The evolution of SALI (solid lines) for a) the chaotic
orbit with initial condition $x_0=0.2$, $y_0=0.2$ and b) the regular
orbit with initial condition $x_0=0.4$, $y_0=0.8$ of the standard
map (\ref{sm:1}) for $K=2$, with respect to the number of iterations
$n$. Note the different scales of the horizontal axis. The Lyapunov
exponent of the chaotic orbit is $L_1 \approx 0.438$. Dashed curves
in panels a) and b) correspond to functions proportional to
$e^{-2L_1 n}$ and $1/n^2$ respectively. It is evident that the
theoretical predictions for the evolution of SALI describe very well
the numerical data. } \label{2D_map_sali_evol}
\end{figure}

In figure \ref{2D_map_sali_evol} we see the different behavior of SALI
for regular and chaotic orbits of the standard map (\ref{sm:1}).
It is exactly this different behavior of the index that allows us to
use SALI for a fast and clear distinction between regions of chaos and
order in the 2--dimensional phase space of the standard map. From the
results of figure \ref{2D_map_sali_evol} and the theoretical
predictions for the evolution of SALI we see that after $n=500$
iterations the value of SALI of a regular orbit becomes of the order
of $10^{-6}$, while for a chaotic orbit SALI has already reached
extremely small values. Thus, the percentage of chaotic orbits for a
given value of $K$ can be computed as follows: We follow the evolution
of orbits whose initial conditions lie on a 2--dimensional grid of
$1000\times 1000$ equally spaced points on the 2--dimensional phase
space of the map (dividing in this way the $(x,y)$ plane in $10^6$
squares) and register for each orbit the value of SALI after $n=500$
iterations. All orbits having values of SALI significantly smaller
than $10^{-6}$ (which correspond to the value SALI reaches after 500
iterations in the case of regular orbits), are characterized as
chaotic. In practice as a good threshold for this distinction we
consider the value $10^{-8}$. Thus, all orbits having $\mbox{SALI}
\leq 10^{-8}$ after $n=500$ iterations are characterized as chaotic,
while all others are considered as non--chaotic.

\begin{figure}[h]
\includegraphics[height=0.3\textheight]{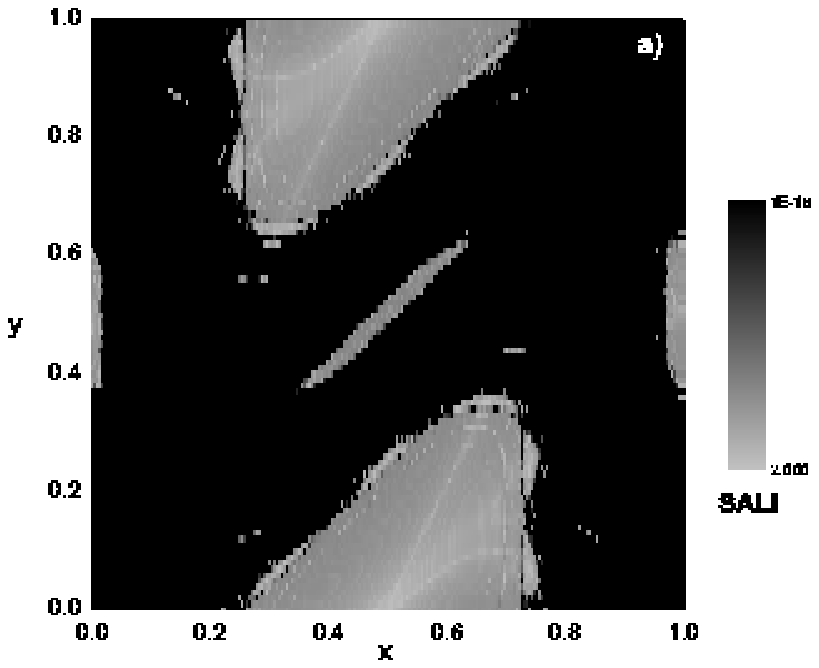}\hspace{0.2cm}
\includegraphics[height=0.3\textheight]{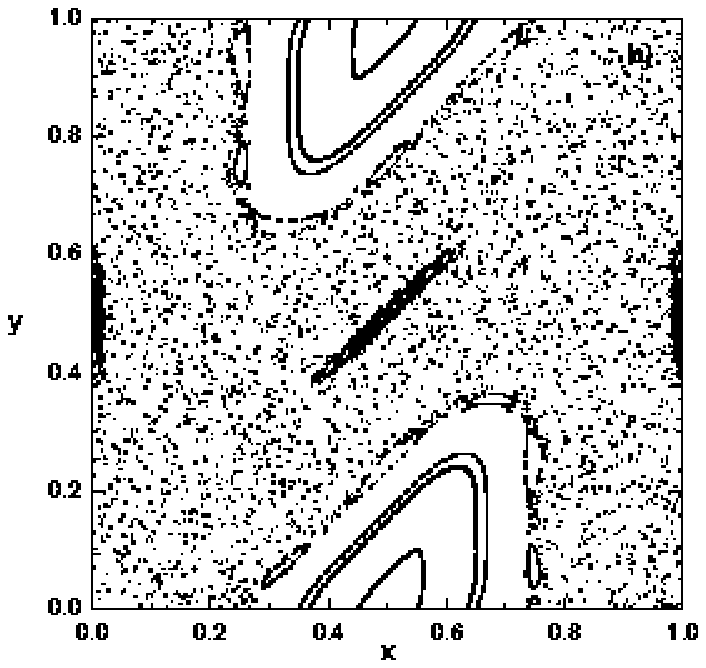}
\caption{a) Regions of different values of SALI after $n=500$
iterations of map (\ref{sm:1}) for $K=2$. b) Phase space portrait of
map (\ref{sm:1}) for the same value of $K$.} \label{2D_map_image}
\end{figure}

In figure \ref{2D_map_image}a) we present the outcome of this
procedure for $K=2$. Each initial condition is colored according to
the color scale seen at the right side of the panel. So, chaotic
orbits, having $\mbox{SALI} \leq 10^{-8}$ are colored black, while
light gray color corresponds to regular orbits having high values of
SALI. Thus, in figure \ref{2D_map_image}a) we can clearly identify
even tiny regions of regular motion which are not easily seen in phase
space portraits of the map (figure \ref{2D_map_image}b)).

\begin{figure}[h]
    \begin{centering}

       \includegraphics[width=7.0cm]{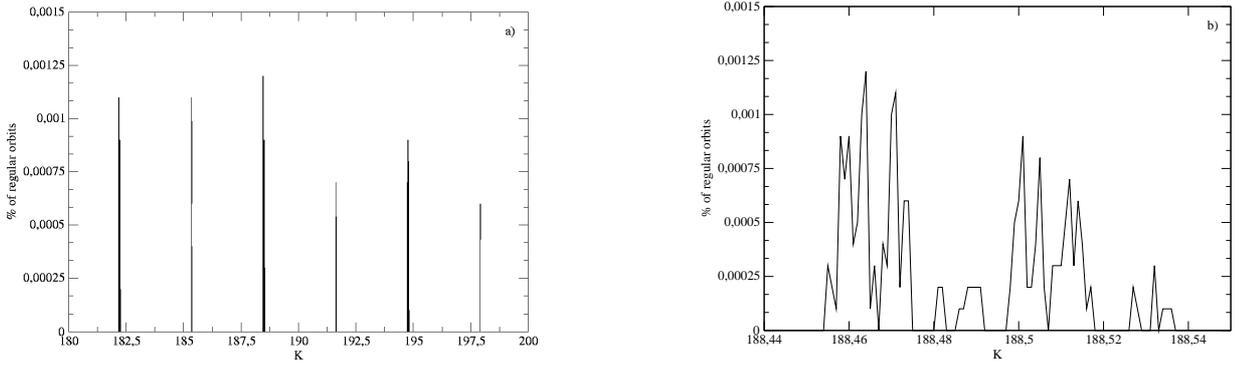}\hspace{2.0cm}
       \includegraphics[width=7.0cm]{win_k_180_200_zoom.eps}
      \end{centering}
        \caption{a) Percentages of regular orbits of map (\ref{sm:1}) as a
        function of the nonlinear parameter $K\in [180,200]$, b) A
        zoom of panel a) in the region of $K\in (188.44,188.55)$.} \label{2D_map_K}
\end{figure}

Using the above--described method we were able to compute very fast
and accurately the percentages of regular motion for large values of
parameter $K$.  In figure \ref{2D_map_K}a) we plot the percentage of
regular orbits for $180\leq K \leq 200$ where $K$ varies with a step
$\delta K = 0.001$. A blow-up of the peak appearing close to $K=188$
is seen in figure \ref{2D_map_K}b). In order to accelerate the
numerical computation we applied the following technique: For each
orbit we compute its SALI value at $n=500$, keeping also track of the
squares on the $(x,y)$ plane that the orbit visits in its
evolution. Then, we attribute the same SALI value to all these
squares. In this way we gain considerably in computational time, since
it is not necessary to perform the same computation for the total
number of the initial conditions. For each value of $K$ a grid of
$1000\times 1000$ initial conditions were used, allowing us to detect
extremely tiny regions of regular motion (note that the percentages of
regular orbits in figure \ref{2D_map_K} remain always less than
$0.0015\%$!).  From the results of figure \ref{2D_map_K} we see a
periodicity of period $2 \pi$ in the appearance of islands of
stability as $K$ varies, in accordance to the results presented in
\cite{Dvo:1}.  In our study we were able to reproduce the results
obtained in \cite{Dvo:1} but with considerably less computational
effort. For example, for $K=2$ instead of using all the $10^{6}$
initial conditions of the $1000\times 1000$ grid, computing the
evolution of only 12425 initial conditions up to $n=500$ iterations
was sufficient for characterizing the total $10^{6}$ points. Thus, for
$K=2$ we were able to compute the percentage of regular orbits on a
$1000\times 1000$ grid mesh by computation only $3 \times 500 \times
12425 \approx 2\cdot 10^7$ iterations of the map (\ref{sm:1}) and its
tangent map, instead for the $5\cdot 10^9$ iterations needed for
obtaining the same result in \cite{Dvo:1}. In particular for the
computation of the data of figure \ref{2D_map_K} we needed only 27
hours of CPU time on an Athlon 64bit, 3.2GHz PC.

\begin{figure}[h!]
\includegraphics[height=0.3\textheight]{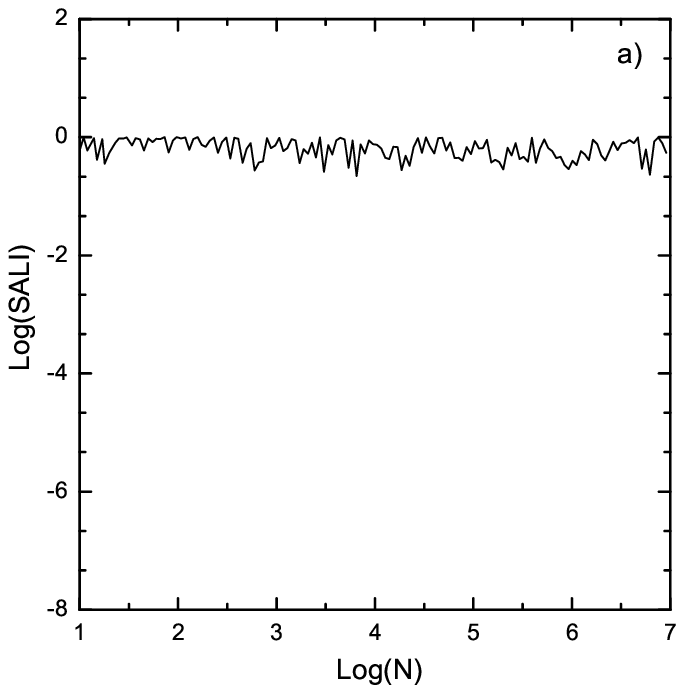}\hspace{0.2cm}
\includegraphics[height=0.3\textheight]{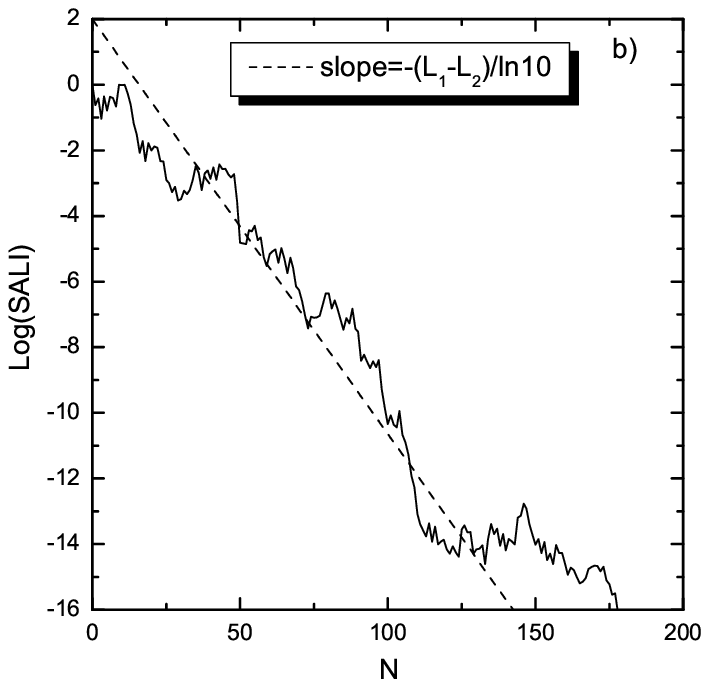}
\caption{The evolution of the SALI for a) the regular orbit with initial condition
$(x_1,x_2,x_3,x_4)=(0.55, 0.1, 0.62, 0.2)$ and b) the chaotic orbit
with initial condition $(x_1,x_2,x_3,x_4)=(0.55, 0.3, 0.62, 0.2)$
of the 4D standard map
with $K=3$ and $\beta=0.1$. We note that the SALI of the chaotic orbit
decays exponentially to zero following the law $\propto
e^{-(L_{1}-L_{2})n}$.} \label{4D_map_orbits}
\end{figure}

A similar study can be also implemented for the 4D standard map,
which is described by the following equations:
\begin{equation}\label{sm4:1}
 \begin{array}{ll}
    x_{1}^{'} = x_1 + x_{2}^{'} & \hbox{} \\
    x_{2}^{'} = x_2 + \frac{K}{2\pi}\sin(2\pi x_1) - \frac{\beta}{\pi}\sin[2\pi (x_3 - x_1)]\\
    x_{3}^{'} = x_3 + x_{4}^{'} & \hbox{} \\
    x_{4}^{'} = x_4 + \frac{K}{2\pi}\sin(2\pi x_3) - \frac{\beta}{\pi}\sin[2\pi (x_1 - x_3)]& \hbox{}
  \end{array}  \,\, (mod\,\,1)\,\, ,
\end{equation}
where $K$ is again the non--linear parameter and $\beta$ the
so--called coupling parameter of the system. In figures
\ref{4D_map_orbits}a),b), we present the evolution of the index both
for regular and chaotic orbits, for $K=3$ and $\beta=0.1$.  For the
regular orbit of figure \ref{4D_map_orbits}a) SALI fluctuates around
to a non--zero value, while for the chaotic orbit of figure
\ref{4D_map_orbits}b) SALI decays exponentially to zero reaching
extremely small values after only a few iterations ($N \approx 150$),
following the exponential law: SALI $\propto e^{-(L_{1}-L_{2})n}$,
with $L_{1}$, $L_{2}$ being the two largest Lyapunov exponents of the
orbit.

We were also able to measure the percentages of chaotic and regular
orbits for the 4D standard map following a similar procedure to the
one used in the case of the 2D map. We considered $10^{6}$ initial
conditions equally spaced in the 4--dimensional phase space of the
system, producing in this way a fine grid of 4--dimensional
hypercubes. Noting that in the case of chaotic orbits only a few
hundreds of iterations are needed for SALI to reach the numerical
accuracy of a computer, i.e.~SALI $\approx 10^{-16}$ (in the case of
the orbits of figure \ref{4D_map_orbits}b) 150 iterations were
sufficient), we started our computation by integrating orbits for only
500 iterations. For each orbit we also kept track of the
4--dimensional hypercubes it visited in its evolution. If the studied
orbit was regular, i.e.~its final SALI value was larger than
$10^{-8}$, its final SALI value was attributed to all the hypercubes
visited by the orbit. If, on the other hand, the orbit was
characterized as chaotic (i.e. its SALI value became $\leq 10^{-8}$),
the evolution of the orbit (but not the evolution of the variational
equations) was extended to 5000 iteration, allowing us to attribute
the computed SALI value to many hypercubes. This procedure decreases
significantly the CPU time needed for the reliable computation of the
percentage of regular motion. In particular for $K=3$ and $\beta=0.1$
the percentage of regular motion was found to be $8,7\%$ after only 1
minute of computations with the same computer used in the 2D case.

\section{Applications to 2D and 3D models of barred galaxies}
\label{Galaxies}

\subsection{The model}
\label{Model_gal}

A 3D rotating  model of a barred galaxy can be described by the Hamiltonian
function:

\begin{equation}\label{eq:Hamilton}
   H=\frac{1}{2} (p_{x}^{2}+p_{y}^{2}+p_{z}^{2})+ V(x,y,z) -
   \Omega_{b} (xp_{y}-yp_{x}).
\end{equation}
The bar rotates around its $z$-axis, while the $x$-axis is along the
major axis
and the $y$-axis is along the intermediate axis. The $p_{x},p_{y}$ and
$p_{z}$ are the canonically conjugate momenta. Finally, $V$ is the
potential, $\Omega_{b}$ represents the pattern speed of the bar and
$H$ is the total energy of the system.

The potential $V$ of our model consists of three components:
\begin{enumerate}
 \item A \textit{disc}, represented by a Miyamoto disc \cite{Miy.1975}:
\begin{equation}\label{Miy_disc}
  V_D=- \frac{GM_{D}}{\sqrt{x^{2}+y^{2}+(A+\sqrt{z^{2}+B^{2}})^{2}}},
\end{equation}
where $M_{D}$ is the total mass of the disc, $A$ and $B$ are the
horizontal and vertical scalelengths, and $G$ is the gravitational
constant.

\item A \textit{bulge} which is modeled by a Plummer sphere whose
potential is:
\begin{equation}\label{Plum_sphere}
    V_S=-\frac{G M_{S}}{\sqrt{x^{2}+y^{2}+z^{2}+\epsilon_{s}^{2}}},
\end{equation}
where $\epsilon_{s}$ is the scalelength of the bulge and $M_{S}$ is
its total mass.

\item A triaxial Ferrers \textit{bar}, the density $\rho(x)$ of which
is:
\begin{eqnarray}\label{Ferr_bar}
  \rho(x)=\begin{cases}\rho_{c}(1-m^{2})^{2} &, m<1  \\
              \qquad 0 &, m\geq1 \end{cases},
\end{eqnarray}
where $\rho_{c}=\frac{105}{32\pi}\frac{G M_{B}}{abc}$ is the
central density, $M_{B}$ is the total mass of the bar and
\begin{equation}\label{Ferr_m}
  m^{2}=\frac{x^{2}}{a^{2}}+\frac{y^{2}}{b^{2}}+\frac{z^{2}}{c^{2}},
\qquad a>b>c> 0,
\end{equation}
with $a,b$ and $c$ being the semi-axes. The corresponding potential is:
\begin{equation}\label{Ferr_pot}
    V_{B}= -\pi Gabc \frac{\rho_{c}}{n+1}\int_{\lambda}^{\infty}
    \frac{du}{\Delta (u)} (1-m^{2}(u))^{n+1},
\end{equation}
where
\begin{equation}\label{mu2}
m^{2}(u)=\frac{x^{2}}{a^{2}+u}+\frac{y^{2}}{b^{2}+u}+\frac{z^{2}}{c^{2}+u},
\end{equation}

\begin{equation}\label{Delta}
\Delta^{2} (u)=({a^{2}+u})({b^{2}+u})({c^{2}+u}),
\end{equation}
$n$ is a positive integer (with $n=2$ for our model) and
$\lambda$ is the unique positive solution of:
\begin{equation}\label{mu2_lamda}
    m^{2}(\lambda)=1,
\end{equation}
outside of the bar ($m\geq 1$), and $\lambda=0$ inside the bar.

The corresponding forces are given analytically in  \cite{Pfe}.
\end{enumerate}

This model has been used extensively for orbital studies $[23\mbox{--}27]$.


\subsection{Numerical results}
\label{Num_gal}

We first applied the SALI method to the 2D bar potential resulting
from the restriction of our study on the $z=p_z=0$ subspace of the
whole phase space of the system. It can be easily seen that, due to
the symmetries of Hamiltonian (\ref{eq:Hamilton}), orbits starting
with $z=p_z=0$ remain for all time on the $(x,y)$ plane. In this case,
the Hamiltonian function governing the motion is derived by setting
$z=p_z=0$ to equations (\ref{eq:Hamilton})--(\ref{Delta}) and the
corresponding Poincar\'{e} Surface of Section (PSS) is 2--dimensional
and can be easily visualized.

\begin{figure*}
  \includegraphics[height=0.325\textheight]{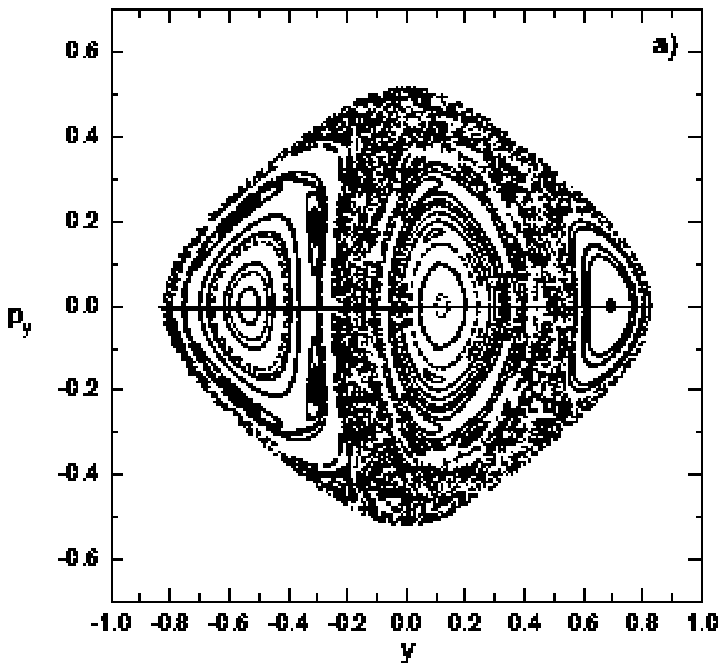}\hspace{0.1cm}
  \includegraphics[height=0.325\textheight]{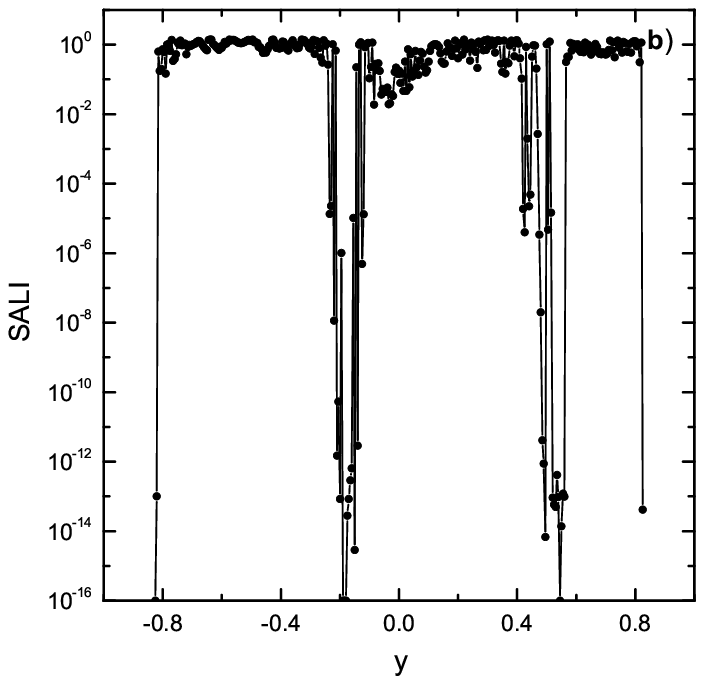}
  \caption{a) Poincar\'{e} surface of section for the $2D$ Ferrers'
  model in $(y,p_{y})$ plane for $H=-0.360$, b) The variation of the
  SALI value for initial conditions chosen on the line $p_{y}=0$ of the
  corresponding PSS of panel a).}
  \label{pss_sali_line}
\end{figure*}

As we have already mentioned, in 2D Hamiltonian systems SALI tends
exponentially to zero for chaotic orbits, while it fluctuates around
a positive number for regular orbits. In figure
\ref{pss_sali_line}a) we present the PSS (plane $(y,p_{y})$) of the
system for $H=-0.360$, which exhibits both regular and chaotic
regions. By choosing orbits with initial conditions on the line
$p_{y}=0$ of the PSS and calculating their SALI values at $t=3000$
we were able to detect very small regions of stability that can not
be visualized easily on the PSS. We plot the corresponding values of
the SALI in figure \ref{pss_sali_line}b). The values of the SALI
tend to zero $(\approx 10^{-16})$, for orbits whose initial
conditions were chosen inside the chaotic regions of the PSS,
contrary to orbits with initials conditions inside the stability
islands whose SALI values retain large positive values.

We also used the SALI method to calculate the percentages of regular
and chaotic motion for initial conditions chosen on the whole plane
$(y,p_y)$. For the value of the Hamiltonian function $E=H=-0.360$ we
tested two sets of initial conditions: set A Having 5000 initial
conditions and set B with 10000 initial conditions. The two set were
used in order to examine the variation of the percentages of the
regular and chaotic orbits for different grids of initial
conditions. We found that for set A the percentage of chaotic orbits
is $22,5\%$, while for set B $28,0\%$ of the orbits were characterized
as chaotic. We note that, as usual, an orbit is considered to be
chaotic if its SALI value becomes $\leq 10^{-8}$.  Thus for this
particular value of the energy, although a finer grid can help us to
detect some small chaotic regions, we can nevertheless derive a good
approximation of the real percentage even with a relatively ``small''
set of initial conditions. Repeating this procedure for several values
of the energy, we were able to follow the change of the fraction of
chaotic and regular orbits in the phase space as the value of $H$
varies.

\begin{figure*}[h]
    \begin{center}
  \includegraphics[height=0.28\textheight]{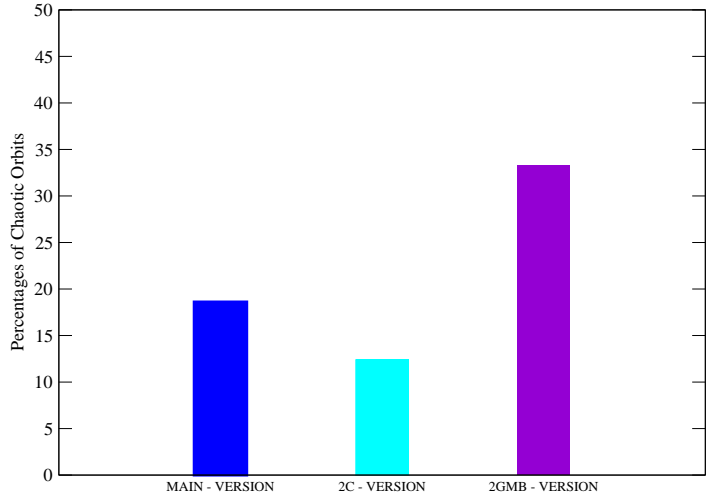}
    \caption{Comparison of the percentages of chaotic orbits
    for our \textit{main} galactic model, described by the Hamiltonian function (\ref{eq:Hamilton})
    with parameters: \ $G=1$, $\Omega_{b}=0.054$,
    $a=6$, $b=1.5$, $c=0.6$, $A=3$, $B=1$, $\epsilon_{s}=0.4$, $M_{B}=0.1$, $M_{S}=0.08$,
    $M_{D}=0.82$, a model with twice the length of the short axis $c=1.2$
    ($\textit{2C}$ - version) and a model with twice the bar mass
    $M_{B}=0.2$ ($\textit{2GMB}$ - version). The system becomes more chaotic as the mass
    of the bar component increases, while a thicker bar results to the
    decrease of chaoticity. The initial conditions in this example are given
    in the $(x,p_{y},z)$ space with $(y,p_{x},p_{z})=(0,0,0)$}.
  \label{perc_chaos_model_a}
    \end{center}
\end{figure*}

We have also studied the complete 3D model, described by the
Hamiltonian (\ref{eq:Hamilton}), for several values of the bar mass
$M_{B}$ and of the semi-minor axis $c$. Our basic model (\textit{main
model}), has the following values of parameters:\ $G=1$,
$\Omega_{b}=0.054$, $a=6$, $b=1.5$, $c=0.6$, $A=3$, $B=1$,
$\epsilon_{s}=0.4$, $M_{B}=0.1$, $M_{S}=0.08$, $M_{D}=0.82$ both for
its 2D and 3D versions. The units used, are: 1 kpc (length), 1 Myr
(time), $2 \times 10^{11}$ solar masses (mass). For each model studied
we considered two sets of initial conditions. The first set contains
orbits with initial conditions in the $(x,p_{y},z)$ space with
$(y,p_{x},p_{z})=(0,0,0)$ and the second one, orbits with initial
conditions in the $(x,p_{y},p_{z})$ space with $(y,z,p_{x})=(0,0,0)$.
Analyzing our numerical results we found that the increase of the bar
mass ($\textit{2GMB}$ - version, with $M_{B}=0.2$) introduces more
chaotic behavior for both sets of initial conditions. In figure
\ref{perc_chaos_model_a} we present the change of percentages of the
chaotic orbits as the parameters of Hamiltonian (\ref{eq:Hamilton})
vary for the first set of initial conditions (similar results where
also found for the second set). The findings are in accordance to the
results obtained in \cite{ABMP} for the 2D case.  On the other hand,
we discovered that when the bar is thicker, i.e.  the length of the
$z$-axis larger ($\textit{2C}$ - version, with $c=1.2$), the system
becomes less chaotic \cite{Man:2}.

Finally, we calculated the percentages of chaotic and regular orbits
for different values of the pattern speed $\Omega_{b}$. From the
orientation of periodic orbits, Contopoulos \cite{Con:4} showed that
bars have to end before corotation, i.e. that $r_{L}>a$, where $r_{L}$
the Lagrangian, or corotation, radius. Comparing the shape of the
observed dust lanes along the leading edges of bars to that of the
shock loci in hydrodynamic simulations of gas flow in barred galaxy
potentials, Athanassoula \cite{Athan:2,Athan:3} was able to set both a
lower and an upper limit to corotation radius, namely $r_{L}=(1.2 \pm
0.2)a$. This restricts the range of possible values of the pattern
speed between a high value that corresponds to the Lagrangian radius
$r_{L}=1.4a$ and a low value that corresponds to $r_{L}=1.0a$. Using
the extremes of this range, we investigated how the pattern speed of
the bar affects the dynamics the system and found that the percentage
of regular orbits is greater in slow bars \cite{Man:3}.

\section{Conclusions}
\label{Conclusions}

In this paper, we used the SALI method of chaos detection to study
the dynamical behavior of 2 and 4 dimensional symplectic maps and of
Hamiltonian models of barred galaxies with 2 and 3 degrees of
freedom. Using the SALI we were able to rapidly identify, in the
phase spaces of the studied systems, even tiny regions of regular
motion and also compute the percentages of regular and chaotic
orbits as the values of the parameters of the systems vary. In the
case of galactic models, we found the influence of some important
physical parameters like the mass, the length of the short $z$-axis
and the pattern speed of the bar, on the chaotic behavior of the
system. In particular, we found that the growth of the mass of the
bar favors the existence of more chaotic orbits, while we observed
that by increasing the length of the short axis of the bar the
percentage of the chaotic orbits decreases.

\acknowledgements{We thank the European Social Fund (ESF), Operational
Program for Educational and Vocational Training II (EPEAEK II), and
particularly the Program PYTHAGORAS II, for funding this work. Visits
between the team members were partially funded by the region
PACA. T.~Manos was supported by the ``Karatheodory" graduate student
fellowship No B395 of the University of Patras. During his visit in
Marseille, his travel and accommodation funds were provided by the
Marie Curie fellowship No HPMT-CT-2001-00338. Ch.~Skokos was supported
by the Marie Curie Intra--European Fellowship No
MEIF--CT--2006--025678. The first author (T.~M.) would also like to
express his gratitude to the Institut de M\'{e}canique C\'{e}leste et
de Calcul des Eph\'{e}m\'{e}rides (IMCCE) of the Observatoire de Paris
for its excellent hospitality during his visit in June 2006, when part
of this work was completed.}

\bibliographystyle{aipproc}   

\IfFileExists{\jobname.bbl}{}
 {\typeout{}
  \typeout{******************************************}
  \typeout{** Please run "bibtex \jobname" to obtain}
  \typeout{** the bibliography and then re-run LaTeX}
  \typeout{** twice to fix the references!}
  \typeout{******************************************}
  \typeout{}
 }

 \vfill
\end{document}